\documentclass[12pt,psamsfonts]{amsart}
\usepackage{amsmath}
\usepackage{amsthm}
\usepackage{amssymb}
\usepackage{amscd}
\usepackage{amsfonts}
\usepackage{amsbsy}
\usepackage{epsfig}

\begin{document}

\title[On the origin of the deflection of light] {On the origin of the deflection of light}

\author[J. Gin\'{e}]
{Jaume Gin\'e}

\address{Departament de Matem\`atica, Universitat de Lleida,
Av. Jaume II, 69. 25001 Lleida, Spain}

\email{gine@eps.udl.es}

\thanks{The author is partially supported by a DGICYT
grant number BFM 2002-04236-C02-01 and by DURSI of Government of
Catalonia ``Distinci\'o de la Generalitat de Catalunya per a la
promoci\'o de la recerca universit\`aria".}

\subjclass{Primary 34C05. Secondary 58F14.}

\keywords{gravitation, retarded systems, functional differential
equations, limit cycle}
\date{}
\dedicatory{}

\maketitle

\begin{abstract}
Action at distance in Newtonian physics is replaced by finite
propagation speeds in classical post--Newtonian physics. As a
result, the differential equations of motion in Newtonian physics
are replaced by functional differential equations, where the delay
associated with the finite propagation speed is taken into
account. Newtonian equations of motion, with post--Newtonian
corrections, are often used to approximate the functional
differential equations. In \cite{G2} a simple atomic model based
on a functional differential equation which reproduces the
quantized Bohr atomic model was presented. The unique assumption
was that the electrodynamic interaction has a finite propagation
speed. In \cite{G3} a simple gravitational model based on a
functional differential equation which gives a gravitational
quantification and an explanation of the modified Titius--Bode law
is described. In \cite{G4} an explanation of the anomalous
precession of Mercury's perihelion is given in terms of a simple
retarded potential, which, at first order, coincides with Gerber's
potential of 1898, and which agrees with the author's previous
works \cite{G2,G3}. In this paper, it is shown how the simple
retarded potential presented in \cite{G4} also gives the correct
value of the gravitational deflection of fast particles of General
Relativity.
\end{abstract}

\section{Introduction}\label{s1}

The history of the deflection of light problem began in 1704, when
Newton proposed in the conclusions of his treatise on Opticks
\cite{IN} the following query:
\begin{quote}
Do not Bodies act upon Light at a distance, and by their action
bend its Rays, and is not this action strongest at the least
distance?
\end{quote}

In fact, this suggestion is not so much radical, because on the
basis of the corpuscular theory of light, and Newton's laws of
mechanics and gravitation, it is easy to conjecture that a ray of
light could be deviated slightly while it passes near a big
massive body, assuming that particles of light respond to
gravitational acceleration similarly to particles of matter.
Johann Georg von Soldner, in 1801, calculated the bending of light
rays grazing the Sun's disk, also referred to in \cite{J}, using
classical mechanics and a hypothetical light velocity of around $3
\cdot 10^5$ km per second. A hundred years later, Einstein
repeated the prediction (in the context of his theories) that
starlight passing by the Sun would be deflected by the Sun's
gravity, so that the apparent distance between stars either side
of the Sun when viewed during an eclipse would be smaller. The
following computation of the deflection of the light, based in the
Newtonian gravitation, is extracted from \cite{M} and \cite{So}
where it is given a complete description of the historical
development of the problem.

For any conical orbit of a small test particle of mass $m$  in a
Newtonian gravitational field around a central mass $M$, the
eccentricity of the unbounded hyperbolic orbit is given by
\[
\varepsilon= \sqrt{1+\frac{2EL^2}{G^2 m^3 M^2}} \ ,
\]
where $E = m v^2/2 - G m M/r$ is the total energy (kinetic plus
potential), $L = m r v_t$ is the angular momentum, $v$ is the
total speed, $v_t$ is the tangential component of the speed, and
$r$ is the radial distance from the center of the mass. Since a
beam of light travels at such a high speed, it will be in a
shallow hyperbolic orbit around an ordinary massive object like
the Sun.

\begin{figure}
\includegraphics[width=7cm]{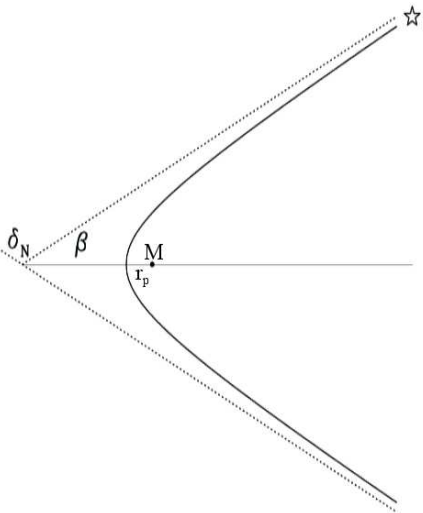}
\caption{\label{fig:hypb}The light ray from a star follows an
unbound hyperbolic orbit about the Sun. For deflection on grazing
incidence, the distance of closest approach is $r_p$.}
\end{figure}

Letting $r_p$ denote the closest approach (the perihelion) of the
beam to the gravitating body, at which $v = v_t$, we have
\[
\varepsilon = \sqrt{1+ \left( \frac{r_p v_t^2}{GM} \right)^2 -
\frac{ 2 r_p v_t^2}{GM} } =1- \frac{r_p v_t^2}{GM} \ ,
\]
As expected, the test particle mass $m$ cancels out above. Now we
set $v_t=c$ (the speed of light) at the perihelion, and from the
geometry of the hyperbola we know that the asymptotes make an
angle of $\beta$ with the axis of symmetry, where $ \cos \beta =
1/ \varepsilon$.

As the hyperbolic orbit shown in Figure 1, the total angular
deflection of the beam of light is $\delta_N=\pi -2 \beta$, which
for small angles $\beta$ and for $M$ much less than $r_p$, is
given in Newtonian mechanics by
\[
\delta_N= \pi- 2 \arccos \left(\frac{1}{\varepsilon}\right) =\pi-
2 \arccos \left( \frac{GM}{GM- c^2r_p} \right) \approx
\frac{2\,GM}{c^2r_p} \ ,
\]

The best natural opportunity to observe this deflection would be
to look at the stars near the perimeter of the Sun during a solar
eclipse.  The mass of the Sun in gravitational units is about $M =
1475$ meters, and a beam of light just skimming past the Sun would
have a closest distance equal to the Sun's radius, $r_p=6.95 \cdot
10^8$ meters.  Therefore, the Newtonian prediction would be
0.000004245 radians, which equals 0.875 seconds of arc.

In 1911, Einstein recaptures the idea of bending the light, see
\cite{E3}. He used the equivalence principle and the equivalent
mass-energy of a photon, together with Special Relativity, to
predict that clocks run at different rates in a gravitational
potential and the bending of light-rays in a gravitational field,
even before he developed the concept of curved-space time. Oddly
enough, the quantitative prediction given in this paper for the
amount of deflection of light passing near a large mass was more
or less identical to the old Newtonian prediction 0.83 seconds of
arc. It wasn't until late in 1915, see \cite{E2}, as he completed
the General Relativity theory, that Einstein realized his earlier
prediction was incorrect, and the angular deflection should
actually be twice the size he predicted in 1911. Only in this
second calculation, published in 1916, where he included the
effect of space-time curvature, he obtained a value
$\delta_{GR}=4GM/ c^2 r_p$.

\section{Deflection of the light from a retarded potential}

Action at distance in Newtonian physics is replaced by finite
propagation speeds in classical post--Newtonian physics. As a
result, the differential equations of motion in Newtonian physics
are replaced by functional differential equations, where the delay
associated with the finite propagation speed is taken into
account. Newtonian equations of motion, with post--Newtonian
corrections, are often used to approximate the functional
differential equations, see, for instance,
\cite{Ch,Ch1,Ch2,Ch3,G1,R,R2}. In \cite{G2} a simple atomic model
based on a functional differential equation which reproduces the
quantized Bohr atomic model was presented. The unique assumption
was that the electrodynamic interaction has finite propagation
speed, which is a consequence of the Relativity theory. A
straightforward consequence of the theory developed in \cite{G2},
and taking into account that gravitational interaction has also a
finite propagation speed, is that the same model is applicable to
the gravitational 2-body problem. In \cite{G3} a simple
gravitational model based on a functional differential equation
which gives a gravitational quantification and an explanation of
the modified Titius--Bode law is described. In \cite{G4} an
explanation of the anomalous precession of Mercury's perihelion is
given in terms of a simple retarded potential, which, at first
order, coincides with Gerber's potential of 1898, and which agrees
with the author's previous works \cite{G2,G3}. In the following,
we show how the values of the anomalous precession of Mercury's
perihelion and the gravitational deflection of fast particles of
General Relativity can be reproduced by a retarded potential.

First, we review the recent intends to solve the problem of the
anomalous precession of Mercury's perihelion and the gravitational
deflection of fast particles of General Relativity in terms of
Weber forces. In \cite{A}, Assis has proposed a theory of
gravitation based on Mach's principle, by postulating that the
resultant force in any body is zero and with a Weber type force
for gravitation. With a suitable coefficient for this force he was
able to reproduce the advance of the perihelion of the planets as
given by General Relativity, fixing in this way the only parameter
of the theory. In \cite{SR} the gravitational deflection of fast
particles was calculated and that the theory does not lead to the
Einstein's value for the deflection of light was shown, see
\cite{E2}. In fact the valued obtained is twice as big. Moreover,
in \cite{SR} it was shown how both results, the advance of
perihelion and fast particle deflection, can be accommodated by
using a modified Weber force. In 1848, Weber presented a velocity
dependent potential
\[
V= \frac{1}{r}\left [ 1 - \frac{\dot{r}^2}{2c^2} \right].
\]
from which the Weber force might be derived. This force form the
basis of the classical Weber's electrodynamics, see \cite{A1}.

The basic equation of motion that it was analyzed in \cite{SR} is
\begin{equation}
\frac{d^2{\bf r}}{dt^2}= -\frac{GM}{r^3} \left [\, 1+
\frac{\xi}{c^2}(r \, \ddot{r}- \alpha \, \dot{r}^2)\right ] {\bf
r} \, , \label{01}
\end{equation}
where ${\bf r}$ is the relative radius vector of the particle with
respect to the Sun. The last term on the right-hand side of
\eqref{01} is called the gravitational Weber force per unit of
mass. With $\alpha=1/2$, \eqref{01} is the expression adopted by
Assis, see \cite{A}, who needed to fix $\xi=6$ to obtain the right
advance of the perihelion of the planets. A law of motion of type
\eqref{01} was first proposed by Tisserand \cite{T} with
$\alpha=1/2$ and $\xi=2$, which corresponds to Weber's law of
electrodynamic action. In \cite{SR}, it is considered the (small)
deflection of a fast particle by the Sun with velocity tending to
the speed of light, and with a distance of closest approach $r_p$
much larger than $GM/c^2$. Moreover, the calculation in Cartesian
coordinates, following the same reasoning of Corinaldesi and
Papapetrou \cite{CP} for the calculation of the deflection of a
spinning fast particle in a Schwarzchild field, was performed.
Choosing the $x$ axis perpendicular to the distance of closest
approach with the $y$ axis along it. Therefore, the deflection
occurs in the plane $xy$ with $dy/dx=0$ at $x=0$. If the
deflection is small $(r_p \gg GM/c^2)$, it is possible to write
$y=r_p$ and $x=c\, t$, in the first approximation. We then have
$r^2=c^2t^2+r_p^2$. Upon integration of $d^2y/dt^2$, with the
condition $dy/dt=0$ at $t=0$, and writing $x=c\, t$, the orbit
equation becomes
\[
\frac{dy}{dt}=-\frac{GM}{r_p \, c^2} \left[ \frac{(1+\xi)
x}{\sqrt{x^2+r_p^2}}-\frac{\xi(\alpha+1)x^3}{3(x^2+r_p^2)^{3/2}}
\right] \, .
\]

The angle of deflection of the path of the particle is $\delta
\approx$ $(dy/dx)_{-\infty}-(dy/dx)_{\infty}$, or
\begin{equation}
\delta = \frac{2GM}{r_p \, c^2}
\left(1+\xi-\frac{\xi(\alpha+1)}{3} \right) \, . \label{03}
\end{equation}
With the value $\alpha=1/2$ adopted by Assis and the value $\xi=6$
needed to reproduce the advance of perihelion, we would have
$\delta=8GM/ r_p \, c^2$, which is twice as big as the value
obtained by Corinaldesi and Papapetrou \cite{CP} for the
deflection of a fast, spinless particle by a Schwartzchild field,
which is also Einstein's value $\delta_{GR}$ for the deflection of
light, see also \cite{W}. To reproduce this value, we need to have
from \eqref{03} the relation
\begin{equation}
\xi(2-\alpha)=3 \, . \label{04}
\end{equation}
Therefore, taking into account that the value of $\xi$ needed to
reproduce the advance of perihelion is $\xi=6$, if we want to
construct a Weber force which also explains the deflection of the
light, equation \eqref{04} gives for $\alpha$ the value of
$\alpha=3/2$, see \cite{SR}.

In \cite{BC}, the Weber-like forces are examined from the point of
view of energy conservation and it is proved that they are
conservative if and only if $\alpha=1/2$. As a consequence, it is
shown that gravitational theories employing Weber-like forces
cannot be conservative and also yield both the precession of the
perihelion of Mercury as well as the gravitational deflection of
light.

\begin{figure}
\includegraphics[width=8cm]{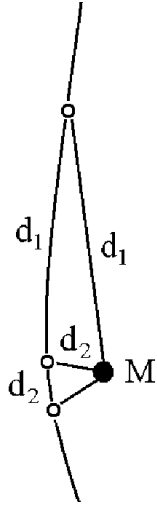}
\caption{\label{fig:hypb}The unbounded hyperbolic orbit of a fast
particle around the Sun with mass $M$.}
\end{figure}

In \cite{G4}, the simple retarded potential was presented:
\begin{equation}
V=- \frac{GMm}{r(t-\tau- \frac{r(t-\tau)}{c})}. \label{05}
\end{equation}
where $r(t-\tau- r(t-\tau)/c)$ is the distance between the masses
when the potential was ``emitted" to go from the emitting particle
to the receiving particle and come back, and $\tau=r(t)/c$. This
retarded potential coincides, at first orders, with Gerber's
potential of 1898 and gives an explanation of the anomalous
precession of Mercury's perihelion because if we develop the
retarded potential \eqref{05} in powers of $\tau$ (up to second
order in $\tau $), then we find that the gravitational force law
associated to this potential is given by
\begin{equation}
f= -\frac{GMm}{r^2}\left(1-\frac{ 3 \dot{r}^2}{c^2}+\frac{ 6
r\ddot{r}}{c^2} \right), \label{06}
\end{equation}
This gravitational force is a Weber type force and it coincides
with the force adopted by Assis \cite{A} (which has the values
$\alpha=1/2$ and $\xi=6$).\\

In the following, we will see that the simple retarded potential
\eqref{05} also gives an explanation of the gravitational
deflection of fast particles. Let $t_0$ be the time when the test
particle is in closest approach (the perihelion), i.e.,
$r(t_0)=r_p$. To evaluate the bending we must take $t \gg t_0$. If
we consider a particle carrying out an unbounded hyperbolic orbit,
at light velocity, and verifying the retarded potential \eqref{05}
we have Figure 2, where
\[
d_1=|{\bf r}(t)|= \int_{t-\tau}^t \sqrt{(x\,'(s))^2+(y\,'(s))^2}
\, ds \, ,
\]
and
\[
d_2=|{\bf r}(t-\tau)|= \int_{t-\tau-r(t-\tau)/c}^{t-\tau}
\sqrt{(x\,'(s))^2+(y\,'(s))^2} \, ds \, ,
\]
where ${\bf r}(t)=(x(t),y(t))$. This fact happens because the
velocity of the particle in the orbit is the light velocity.
Therefore, from Figure 2, we have that $r(t-\tau) \ll r(t)$ and
hence, in this case
\begin{equation}
V=- \frac{GMm}{r(t-\tau- \frac{r(t-\tau)}{c})} = - \frac{GMm}{r(t-
\frac{r(t)+r(t-\tau)}{c})} \approx - \frac{GMm}{r(t-\tau)}.
\label{07}
\end{equation}
If we develop the approximation of the retarded potential
\eqref{07} in powers of $\tau$ (up to second order in $\tau $), we
obtain
\begin{equation}
V \approx -\frac{GMm}{r} \left [ 1 + \frac{\dot{r}}{r} \ \tau +
\left( \frac{\dot{r}^2}{r^2}- \frac{\ddot{r}}{2r} \right) \ \tau^2
\right], \label{eqs5}
\end{equation}
To develop some easier calculations we can reject, on the right
hand side of expression \eqref{eqs5}, the term with $\ddot{r}$ (in
fact this term is negligible and it only gives rise to terms of
higher order). Hence, at this approximation, we obtain the
velocity--dependent potential
\begin{equation}
V \approx -\frac{GMm}{r} \left [ 1 + \frac{ \dot{r}}{r} \ \tau +
\frac{\dot{r}^2}{r^2} \ \tau^2 \right], \label{eqs6}
\end{equation}
Introducing the expression of the delay $\tau=r(t)/c$ in
\eqref{eqs6} we have:
\begin{equation}
V \approx -\frac{GMm}{r} \left [ 1 + \frac{ \dot{r}}{ c } +
 \frac{ \dot{r}^2}{ c^2 } \right]. \label{eqs7}
\end{equation}
On this basis, of this velocity-dependent potential function
\eqref{eqs7}, the gravitational force law is given by substituting
the potential function \eqref{eqs7} into the equation:
\begin{equation}
f=\frac{d}{dt} \left( \frac{\partial V}{\partial \dot{r}} \right)-
\frac{\partial V}{\partial r} = -\frac{GMm}{r^2}\left(1-\frac{
\dot{r}^2}{c^2}+\frac{ 2 r\ddot{r}}{c^2} \right). \label{eqs8}
\end{equation}
Hence, we obtain a Weber type force with $\alpha=1/2$ and $\xi=2$.
In fact, the force described in \eqref{eqs8} coincides with the
Weber type force proposed by Tisserand \cite{T}, which corresponds
to Weber's law of electrodynamic action. It is easy to see that
these values verify equation \eqref{04} and hence the
gravitational force law \eqref{eqs8} gives the correct value of
the gravitational deflection of fast particles of General
Relativity.

\section{Concluding remarks}

We have seen how the values of the advance of the perihelion of
the planets and the gravitational deflection of the fast particles
of General Relativity can be reproduced with the simple retarded
potential \eqref{05}. In the first case, we directly develop the
retarded potential in powers of $\tau$. In the second case, first
we impose that the velocity of the particle is close to the light
velocity, which implies $r(t-\tau) \ll r(t)$ and after we develop
the approximation of the retarded potential in powers of $\tau$.
It should also mentioned that, although this model gives identical
results as the one obtained by using General Relativity, these
theories are based on different concepts and mathematical tools.
In addition, the differences appear in the terms of higher order
in $\tau$, and the improvement of observations and experimental
techniques will accomplish a good test for both theories in the
future.

In \cite{A}, Assis proposes the postulate that the resultant force
acting on any body is zero. With this postulate and the Weber
force law \eqref{06} for gravitation, he obtains equations of
motion and concludes that all inertial forces are due to
gravitational interaction with other bodies in the universe, as it
was suggested by Mach. All these arguments are accomplished in a
strictly relational theory, see also \cite{A2}. Now, we have an
important framework which explains the introduction in these
models of the Weber's force. These forces are, in fact,
approximations of retarded forces, taking into account the finite
propagation speed. A coherent theory of the inertia according with
the Mach's principle lacks to be given. We hope to give an answer
in a future
work.\\

\noindent{\bf Acknowledgements:}

The author would like to thank Prof. H. Giacomini from
Universit\'e de Tours and Prof. M. Grau from Universitat de Lleida
for several useful conversations and remarks.


\begin{thebibliography}{99}

\bibitem{A} {\sc A.K.T. Assis}, {\it On Mach's principle}, Found. Phys. Lett. {\bf 2} (1989), 301--318.

\bibitem{A1} {\sc A.K.T. Assis}, {\it Weber's electrodynamics}, Kluwer Academic Publishers, Dordrecht, 1994.

\bibitem{A2} {\sc A.K.T. Assis}, {\it Relational Mechanics}, Apeiron, Montreal, 1999.

\bibitem{BC} {\sc F. Bunchaft \& S. Carneiro}, {\it Weber-like interactions and energy
conservation}, Found. Phys. Lett. {\bf 10} (1997) 393--401

\bibitem {Ch} {\sc C. Chicone}, {\it What are the equations of
motion of classical physics?}, Can. Appl. Math. Q. {\bf 10}
(2002), no. 1, 15--32.

\bibitem {Ch1} {\sc C. Chicone, S.M. Kopeikin, B. Mashhoon and D. Retzloff},
{\it Delay equations and radiation damping}, Phys. Letters {\bf A
285} (2000), 17--16.

\bibitem {Ch2} {\sc C. Chicone}, {\it Inertial and slow manifolds for
delay equations with small delays}, J. Differential Equations {\bf
190} (2003), no. 2, 364--406.

\bibitem {Ch3} {\sc C. Chicone}, {\it Inertial flows, slow flows, and
combinatorial identities for delay equations}, J. Dynam.
Differential Equations {\bf 16} (2004), no. 3, 805--831.

\bibitem {CP} {\sc E. Corinaldesi \& A. Papapetrou}, {\it Spinning test
particles in General Relativity 1.- 2.} Proc. Roy. Soc. Lond. {\bf
A209} (1951) 248--268.

\bibitem{E3} {\sc A. Einstein}, {\it \"Uber den Einflu\ss~der Schwerkraft
auf die Ausbreitung des Lichtes (On the influence of gravitation
on the propagation of light)} Ann. Phys. {\bf 35} (1911),
898--908.

\bibitem {E1} {\sc A. Einstein}, {\it Erkl\"arung der Perihelbewegung
des Merkur aus der allgemeinen Relativit\"atstheorie (Explanation
of the perihelion motion of mercury from the general theory of
relativity)}. K\"oniglich Preu\'aische Akademie der
Wissenschaften, Sizungsberichte (1915), 831--839.

\bibitem{E2} {\sc A. Einstein}, {\it Die Grundlage der allgemeinen
Relativit\"{a}etstheorie (The foundation of the general theory of
relativity)}. Ann. Phys. {\bf 49} (1916), 769--822.

\bibitem{Ge1} {\sc P. Gerber}, {\it Die r\"aumliche und zeitliche
Ausbreitung der Gravitation (Space and temporary propagation of
gravitation).} Z. Math. Phys. {\bf 43} (1898), 93--104.

\bibitem{Ge2} {\sc P. Gerber}, {\it Die Fortpflanzungsgeschwindigkeit der
Gravitation (The propa\-gation-velocity of gravitation)}. Ann.
Phys. {\bf 52} (1917), 415--444.

\bibitem{G1} {\sc J. Gin\'e}, {\it \ On the classical descriptions of the
quantum phenomena in the harmonic oscillator and in a charged
particle under the coulomb force}, Chaos Solitons Fractals {\bf
26} (2005), 1259--1266.

\bibitem{G2} {\sc J. Gin\'e}, {\it \ On the origin of quantum
mechanics}, physics/0505181, preprint, Universitat de Lleida,
2005.

\bibitem{G3} {\sc J. Gin\'e}, {\it \ On the origin of gravitational
quantization: the Titius--Bode law}, physics/0507072, preprint,
Universitat de Lleida, 2005.

\bibitem{G4} {\sc J. Gin\'e}, {\it \ On the origin of the anomalous
precession of Mercury's perihelion}, physics/0510086, preprint,
Universitat de Lleida, 2005.

\bibitem{J} {\sc S.L. Jaki}, {\it \ Soldner, Johann, Georg, Von and
the gravitational bending of light, with an English translation of
his essay on it published in 1801}, Found. Phys. {\bf 8} (1978),
927--950.

\bibitem{M} {\sc MathPages} http://www.mathpages.com/rr/s6-03/6-03.htm

\bibitem{IN} {\sc I. Newton}, {\it Opticks. A Treatise of the Reflections,
Refractions, Inflections \& Colours of Light}, Dover, New York,
1979.

\bibitem{P1} {\sc H. Poincar\'e}, {\it M\'emoire sur les courbes
d\'efinies par les \'equations diff\'erentielles.} Journal de
Math\'ematiques {\bf 37} (1881), 375-422; {\bf 8} (1882), 251-296;
Oeuvres de Henri Poincar\'e, vol. I, Gauthier-Villars, Paris,
(1951), pp. 3-84.

\bibitem{SR} {\sc S. Ragusa}, {\it Gravitation with modified weber
force}, Found. Phys. Lett. {\bf 5} (1992), 585--589.

\bibitem{R} {\sc C.K. Raju}, {\it The electrodymamic 2-body problem
and the origin of quantum mechanics}, Found. Phys. {\bf 34}
(2004), 937--962.

\bibitem{R2} {\sc C.K. Raju}, {\it Time: towards a consistent theory},
Kluwer academic, Dordrecht, 1994.

\bibitem{So} {\sc D.S.L. Soares}, {\it \ Newtonian gravitational
deflection of light revisited}, physics/0508030, preprint, 2005.

\bibitem{T} {\sc F. Tisserand}, {\it \ Sur le mouvement des
Plan\`etes autour du Soleil, d'apr\`es ña loi \'electrodynamique
de Weber}, Compt. Rend. Acad. Sci. (Paris) {\bf 75}, 760 (1872);
{\bf 110}, 313 (1890).

\bibitem{W} {\sc S. Weinberg}, {\it Gravitation and Cosmology}, Wiley, New
York, 1972, p. 188.

\end{thebibliography}
\end{document}